\documentclass[aps,epsfig]{revtex4}
\usepackage[dvips]{graphicx}
\begin{document}
\title{Folding model analysis of proton scattering from $^{18,20,22}$O nuclei}
\author{D. Gupta$\footnote{E-mail:dhruba@ipno.in2p3.fr}$, E. Khan, Y. Blumenfeld}
\address{Institut de Physique Nucleaire, IN2P3-CNRS, Universite de Paris-Sud, 91406 Orsay Cedex, France}
\date{\today}
\begin{abstract}
The elastic and inelastic proton scattering on $^{18,20,22}$O nuclei
are studied in a folding model formalism of nucleon-nucleus optical
potential and inelastic form factor. The DDM3Y effective interaction
is used and the ground state densities are obtained in continuum
Skyrme-HFB approach. A semi-microscopic approach of collective form
factors is done to extract the deformation parameters from inelastic
scattering analysis while the microscopic approach uses the
continuum QRPA form factors. Implications of the values of the
deformation parameters, neutron and proton transition moments for
the nuclei are discussed. The p-analyzing powers on $^{18,20,22}$O
nuclei are also predicted in the same framework.

\vskip .2 true cm \noindent Keywords: Elastic and Inelastic Proton
Scattering; Analyzing Power; Effective Interaction; Folding Model; DDM3Y; HFB; QRPA
\end{abstract}
\maketitle

\section {Introduction}
The study of the oxygen isotopic chain deserves special attention
since the neutron drip line was shown to be located at A =
24~\cite{TA97,SA99}. Thus rapid structural changes are expected as
we move from $^{16}$O towards its neutron-rich exotic isotopes.
Several theoretical and experimental endeavors also indicate N = 14
and N = 16 shell closures~\cite{BE06}. In addition, there is
opportunity to track neutron and proton contributions of the 2$^+$
state as the neutron drip line is approached~\cite{JE99}. Proton
scattering is widely used as a means to study both macroscopic and
microscopic aspects of nuclear structure~\cite{AM00,GU02,GU05,KH03}.
A suitable realistic effective nucleon-nucleon (NN) interaction is
also needed in the analysis~\cite{GU05}. A folding model approach is
followed in this study of proton scattering from $^{18,20,22}$O at
43, 43, 46.6A MeV respectively, measured at GANIL~\cite{KH00,BE06}.
The earlier p-$^{18}$O differential cross section and analyzing
powers for the ground state and 2$_1^+$ level at 24.5A MeV incident
energy~\cite{ES74} are also included in the analysis. The folding
model, which relates the density profile of the nucleus with the
scattering cross sections is powerful tool for analyzing
nucleus-nucleus scattering data at a few tens of
MeV/nucleon~\cite{GU05,GU02,KH00,KH03,ES74,BE06,GU00,PE93,RE74}.
Thus it is very appropriate for studying nuclei with extended
density distributions. It should be noted that the same formalism is
being followed to provide description of radioactivity, $\alpha$ and
heavy ion scattering in a double folding model as well as nuclear
matter and p-elastic/inelastic scattering in a single folding
model~\cite{BA04,GU05}

\section {Theoretical formulation}
In a single folding calculation the nucleon-nucleus potential is
obtained by using the density distribution of the nucleus and the
nucleon-nucleon effective interaction~\cite{SA79} as,
\begin{equation}
U(\vec{r_1}) = \int \rho_2(\vec{r_2}) v(|\vec{r_1}-\vec{r_2}|)d^3\vec{r_2}\\
\end{equation}
where $\rho_2(\vec{r_2})$ is density of the nucleus at $\vec{r_2}$
and $v(r)$ is the effective interaction between two nucleons at the
sites $\vec{r_1}$ and $\vec{r_2}$. The finite range M3Y effective
interaction $v(r)$~\cite{r9}, is based upon a realistic G-matrix and
was constructed in an oscillator basis. Effectively it is an average
over a range of nuclear densities as well as energies and thereby
has no explicit dependence on density or energy. The only rather
weak energy dependent effect is contained in an approximate
treatment of single-nucleon knock-on exchange. At lower energies,
the density and energy averages are adequate for the real part of
the heavy ion optical potentials. For scattering at higher energies,
explicit density dependence was introduced~\cite{r10,r11}. The
present calculations use this density dependent M3Y (DDM3Y)
effective NN interaction with an added zero-range pseudo potential
given by,

\begin{equation}
  v(r,\rho,E) = t^{\rm M3Y}(r,E)g(\rho,E)
\end{equation}
\noindent where $E$ is incident energy and

\begin{equation}
  t^{\rm M3Y} = 7999 \frac{e^{ - 4r}}{4r} - 2134\frac{e^{- 2.5r}}{2.5r} + J_{00}(E) \delta(r)
\end{equation}
\noindent The zero-range pseudo-potential~\cite{r10} represents the
single-nucleon exchange term and is given by

\begin{equation}
 J_{00}(E) = -276 (1 - 0.005E / A) {\rm MeV.fm^3}
\end{equation}
\noindent while the density dependent part is taken to be \cite{r11}

\begin{equation}
g(\rho, E) = c (1 - b(E)\rho^{2/3})
\end{equation}
\noindent taking care of the higher order exchange and Pauli
blocking effects. Here $\rho~=\rho_2$ is the spherical ground state density
of the nucleus. The constants of this interaction $c$ and $b$ when
used in single folding model description, are determined by nuclear
matter calculations~\cite{BA04} as 2.07 and 1.62 fm$^2$
respectively.

\section {Calculation and Analysis}

The neutron and proton matter densities were calculated in the
continuum Skyrme-Hartree-Fock-Bogoliubov (HFB) approach and the
microscopic transition densities were obtained within the framework
of the  continuum QRPA (quasi-particle random phase approximation)
formalism~\cite{KH00,KH02,BE06}. Fig. 1 shows the HFB ground state
densities and QRPA transition densities for the 2$_1^+$ states of
$^{18,20,22}$O. In both HFB and QRPA calculations, all quasiparticle 
states below 60 MeV are considered, allowing to take into account 
seven oscillator shells.

In the present analysis both the real ($V$) and volume imaginary
($W$) parts of the folded nuclear potentials are assumed to have the
same shape~\cite{GU05}, i.e. $V_{\rm micro}(r)~=~V + iW$ = ($N_{\rm
R}$ + $iN_{\rm I}$)$U$($r_1$) where, $N_{\rm R}$ and $N_{\rm I}$ are
the renormalization factors for real and imaginary parts
respectively, obtained by fitting the elastic scattering data. In
addition to real and volume imaginary folded potentials, best fit
phenomenological surface imaginary and spin-orbit potentials 
(4$W_D$= 23.08, $r_I$= 1.17, $a_I$= 0.69, 4$V_{s.o}$= 23.60, 
$r_{s.o}$= 0.88, $a_{s.o}$= 0.63) are also used as in~\cite{GU05}. 
The parameters are same for all targets, with volume integral 
per nucleon of the surface imaginary potential for $^{18,20,22}$O 
as 176.5, 168.8 and 162.2 MeV fm$^{2}$ respectively. In~\cite{KH00}, 
the first diffraction
minimum of the elastic scattering on $^{20}$O around $\theta_{\rm
cm}$ = 35$^o$ shows a discrepancy of about 50$\%$ with both
phenomenological and microscopic calculations. It was thought that
improvement of the imaginary part of the optical potential is
required for these unstable nuclei. The more recent work~\cite{KH03}
as well as the present formalism using a DDM3Y interaction
satisfactorily explains the data.

The nucleon-nucleus optical potentials from elastic scattering best
fits are subsequently used to generate the distorted waves for
inelastic scattering amplitude calculations in DWBA formalism (Fig.
2). The calculations are performed using the code
DWUCK4~\cite{DWUCK4}. Initially, the conventional approach of
collective vibrational model ($\beta \frac{dV}{dr}$) is used to
obtain the transition form factors. The deformation parameters
$\beta$ are determined by fitting the inelastic scattering angular
distributions. Table 1 gives the renormalization factors, $\beta$
values, $\chi^2$/N  for the folding model analysis. Compared to
earlier results~\cite{KH00}, the present $\beta$ value of 0.33 for
$^{18}$O 2$_1^+$ agrees well while giving a slightly lower value of
0.46 for $^{20}$O 2$_1^+$ state. The high $\beta$ value of 0.50 for
$^{20}$O 2$_1^+$ state was explained~\cite{JE99} as neutrons playing
a disproportionately large role in the excitation. The present
calculations give the lowest $\beta$ value of 0.26 for $^{22}$O.

A more microscopic approach is useful to investigate low-lying
excitations in exotic nuclei~\cite{BE06}. Calculations have also been
performed (Fig. 2) with form factors using QRPA transition
densities. It may be noted that two probes namely (p,p') and $B(E2)$
are required to derive proton and neutron contributions.
The adopted values~\cite{RA01} for the reduced quadrupole
transition rate $B(E2)$ from the ground state to 2$_1^+$ state for
$^{18,20,22}$O are used in this work and given in Table 2. Though
the energy of the 2$_1^+$ state in $^{20}$O is lower than $^{18}$O,
the $B(E2)$ value is considerably lower indicating a lesser degree 
of collectivity for $^{20}$O~\cite{TH00}. 

In this context, the neutron/proton matrix element is defined as,
$M_{n,p}=\int\delta\rho_{n,p}^{(\nu)}(r)r^{l+2}dr$,
$\delta\rho_{n,p}^{(\nu)}(r)$  is the neutron (proton) transition
density between ground state and excited state $|\nu>$ and $B(E2) =
M_p^2$. As in~\cite{KH01}, the proton transition density is first
normalized (if required) to match with the experimental $B(E2)$
value. Then the magnitude of the neutron transition density is
adjusted to give best fits to the inelastic scattering data. In an
earlier folding model approach~\cite{KH03} substantially high values
of $M_n/M_p$ of 1.80 and 4.25 for $^{18}$O and $^{20}$O respectively
were reported. The results of the present work are shown in Table 2
along with experimental $M_n/M_p$ value~\cite{BE06,KH00}. Though the
$B(E2)$ value is lower, the present results on $M_n/M_p$ of $^{20}$O
confirms its strong neutron contribution. Thus, for both
$^{18}$O and $^{22}$O the $M_n/M_p$ (0$^+$ $\rightarrow$ 2$^+$)
values of 1.12 and 2.00 are close to the $N/Z$ ($N$ and $Z$ are the
neutron and proton numbers) values of 1.25 and 1.75 while for
$^{20}$O the $M_n/M_p$ (0$^+$ $\rightarrow$ 2$^+$) value of 3.34 is
high as compared to the $N/Z$ value of 1.50.

In Fig. 3, the elastic cross sections for p + $^{18,20,22}$O
scattering have been plotted as ratios to the Rutherford cross 
sections (scaled by a factor of 10 for clarity). Using the code 
CHUCK3~\cite{DWUCK4}, the coupled channel (CC) calculations of the 
inelastic scattering cross sections are also shown in Fig 3. Only 
ground state and the first excited state have been considered and 
coupling is both ways. The DWBA and CC calculations using same form 
factors are nearly identical as can also be ascertained by the  
moderately high values of $N_{\rm R}$ from elastic scattering. Using 
different collective form factors give slightly different results. 
The dotted lines in Fig. 3 correspond to CC calculations using 
present collective form factors. The dashed lines in Fig. 3 
correspond to form factors obtained from Legendre expansion of 
volume and surface Wood-Saxon potential, where monopole part of the 
potential is extracted.

It may be noted that the spin-orbit coupling in the present study is
treated phenomenologically from the best fit optical potential
parameters, since a phenomenological Thomas form for the spin-orbit
potential gives a good description of the elastic data while the
target excitation is simply represented by a nuclear transition
density. The central transition potential explains the inelastic
cross section reasonably well, giving a dominant contribution while
the transition spin-orbit potential has only a minor
effect~\cite{KH002}. The vector analyzing powers for $^{18}$O(p,p')
scattering at E= 24.5A MeV~\cite{ES74} are also compared with the
calculations and the results are shown in Fig.~4. The Fig. 5 gives
the predictions of analyzing powers for proton scattering from
$^{18,20,22}$O nuclei at 43, 43 and 46.6A MeV respectively, both for
the ground state as well as the 2$_1^+$ excited states. The
calculations are done on the same footing as the cross sections with
HFB and QRPA approaches. For the ground state, while the 43A MeV p +
$^{18,20}$O angular distributions are nearly same, the p + $^{22}$O
analyzing power at slightly different energy of 46.6A MeV is
marginally different. For the excited state all of them are nearly
identical. It would indeed be interesting to observe how well the
calculations agree with any future experimental data on p +
$^{18,20,22}$O analyzing powers.

\section {Conclusion}

In this work, a single folding model approach is followed to study
the cross sections as well as analyzing powers of proton scattering
from neutron rich oxygen isotopes. In the DWBA formalism, both
collective-model (deforming the optical potential) as well as
microscopic calculations using HFB and QRPA densities with the DDM3Y
effective interaction are carried out. The present calculations
satisfactorily reproduce the experimental data showing a high
neutron contribution to the excited state and a large $\beta$ value
for $^{20}$O as compared to $^{18,22}$O. Coupled channel calculations
of the inelastic scattering cross sections reflect insignificant
coupling. Predictions of the analyzing powers for ground state and 
excited state for proton scattering on the Oxygen isotopes at the 
same energies are given. With the availability of a larger data set 
on unstable nuclei, similar microscopic studies would be immensely 
helpful for nuclear structure and reaction studies.

\begin{table}
{\bf Table 1:} \\
Renormalizations of DDM3Y folded potentials for p + $^{18,20,22}$O
scattering at incident energy ($E/A$) and excited state energy
($E^*$) in MeV, angular momentum transfer ($l$), deformation
parameter ($\beta$), volume integral ($J/A$) of the real folded
potential in MeV fm$^3$ and $\chi^2$/N values from
best-fits to the elastic and inelastic scattering data\\
\setlength{\tabcolsep}{2.0 mm}
\begin{tabular}{cccccccccc}
\hline \hline \multicolumn{1}{c}{Nucleus}&
\multicolumn{1}{c}{$E/A$}& \multicolumn{1}{c}{$E^*$}&
\multicolumn{1}{c}{$N_{\rm R}$}& \multicolumn{1}{c}{$N_{\rm I}$}&
\multicolumn{1}{r}{$l$}& \multicolumn{1}{c}{$\beta$}&
\multicolumn{1}{r}{$\chi^2_{\rm el}$/N}&
\multicolumn{1}{r}{$\chi^2_{\rm inel}$/N}&
\multicolumn{1}{r}{$J/A$}\\
\hline
$^{18}$O$^*$&43.0&1.98&0.90&0.08&2&0.33&7.55&0.81&-470.3\\
$^{20}$O$^*$&43.0&1.67&0.88&0.08&2&0.46&2.04&2.15&-457.6\\
$^{22}$O$^*$&46.6&3.17&0.73&0.08&2&0.26&5.05&4.61&-371.8\\
\hline\hline
\end{tabular}
\end{table}

\begin{table}
{\bf Table 2:} \\
Reduced electric quadrupole transition probability $B(E2)$ in
e$^2$fm$^4$, ratio of neutron to proton transition moment
$M_n$/$M_p$. The proton part of the QRPA transition density was
scaled to give experimental transition probability, while the
neutron part was adjusted by the
best fit to the inelastic scattering data\\

\setlength{\tabcolsep}{2.0 mm}
\begin{tabular}{ccccccccc}
\hline \hline \multicolumn{1}{c}{Nucleus}&
\multicolumn{1}{c}{$N/Z$}& \multicolumn{1}{c}{$E^*$}&
\multicolumn{1}{c}{$B(E2)$}& \multicolumn{1}{c}{$E/A$}&
\multicolumn{1}{r}{$\chi^2_{\rm inel}$/N}&
\multicolumn{1}{c}{($M_n$/$M_p$)$_{expt}$}&
\multicolumn{1}{c}{$M_n$/$M_p$}&
\multicolumn{1}{c}{($M_n$/$M_p$)/($N/Z$)}\\
\hline
$^{18}$O$^*$&1.25&1.98&45.1$\pm$2.0&43.0&2.56&1.10$\pm$0.24&1.12&0.90\\
$^{20}$O$^*$&1.50&1.67&28.1$\pm$2.0&43.0&6.59&3.25$\pm$0.80&3.34&2.23\\
$^{22}$O$^*$&1.75&3.17&21.0$\pm$8.0&46.6&0.60&2.50$\pm$1.00&2.00&1.14\\
\hline\hline
\end{tabular}
\end{table}

\begin{figure}[h]
\includegraphics[width = 0.70\hsize,clip]{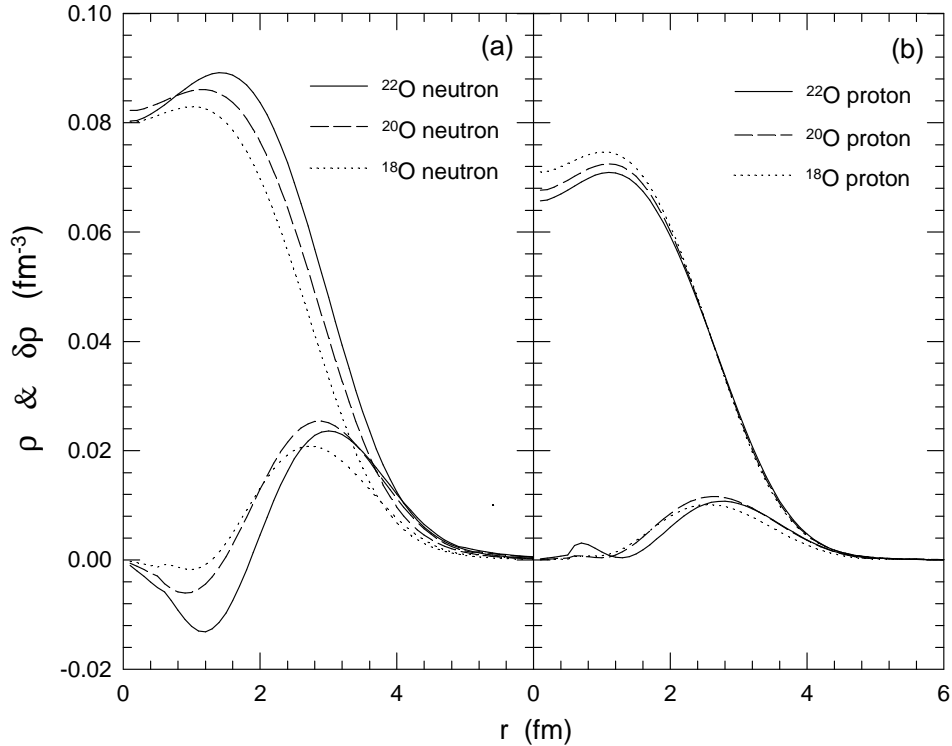}
\caption{(a) Neutron ground state densities (continuum Skyrme-HFB)
and neutron transition densities (QRPA) for the 2$_1^+$ states of
$^{18,20,22}$O, (b) same as (a) but for protons}
\end{figure}

\begin{figure}[h]
\includegraphics[width = 0.70\hsize,clip]{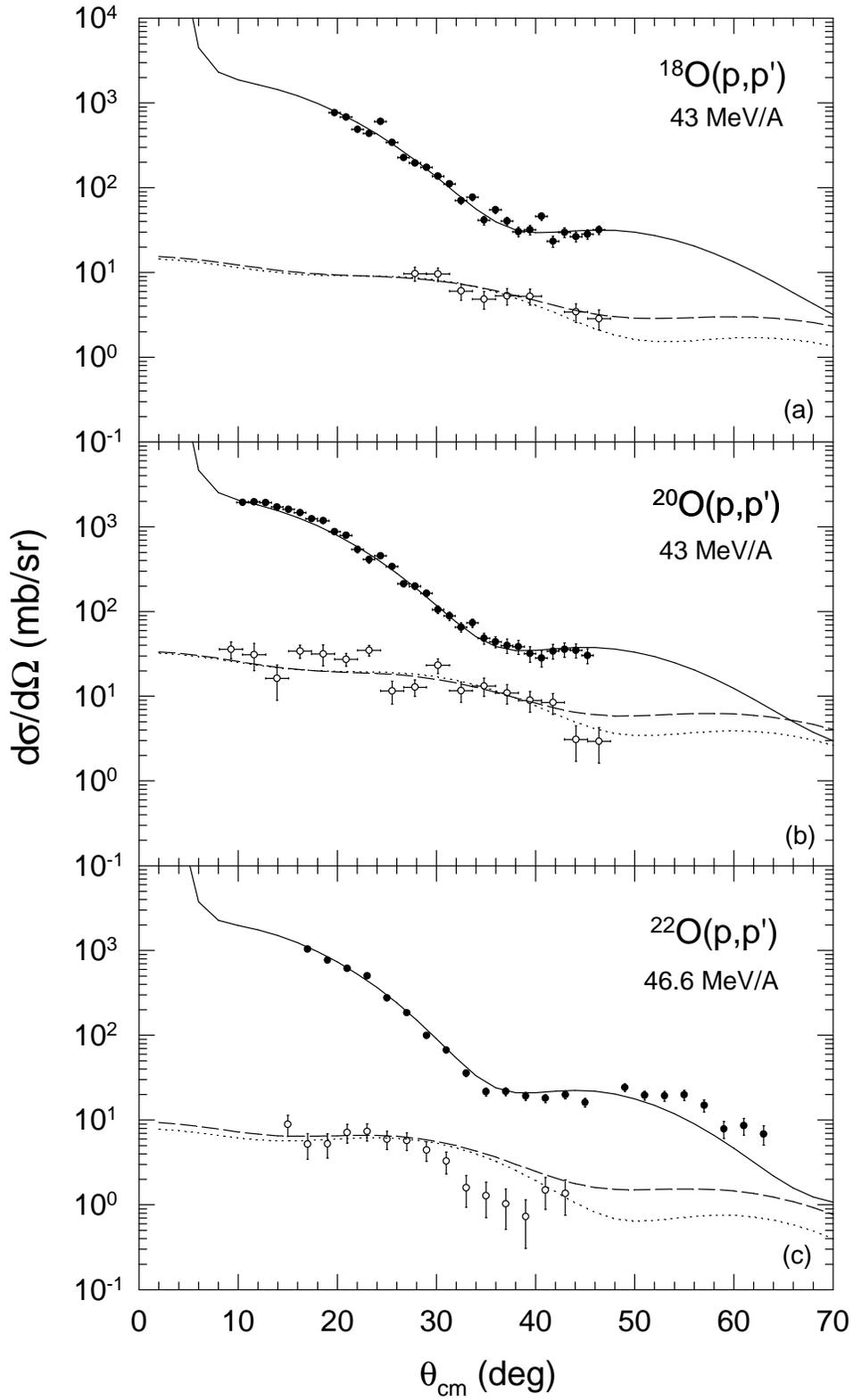}
\caption {The experimental angular distributions and folding model
calculations of (a) p + $^{18}$O at 43A MeV for elastic and
inelastic [$E^*$ = 1.98 MeV (2$^+$)] scattering~\cite{KH00}, (b) p +
$^{20}$O at 43A MeV for elastic and inelastic [$E^*$ = 1.67 MeV
(2$^+$)] scattering~\cite{KH00}, (c) p + $^{22}$O at 46.6A MeV for
elastic and inelastic [$E^*$ = 3.17 MeV (2$^+$)]
scattering~\cite{BE06}. The corresponding $N_{\rm R}$, $N_{\rm I}$
values are given in Table 1. The continuous, dashed (QRPA), dotted
(collective) lines correspond to calculations for elastic and
inelastic cross sections respectively.}
\end{figure}

\begin{figure}[h]
\includegraphics[width = 0.70\hsize,clip]{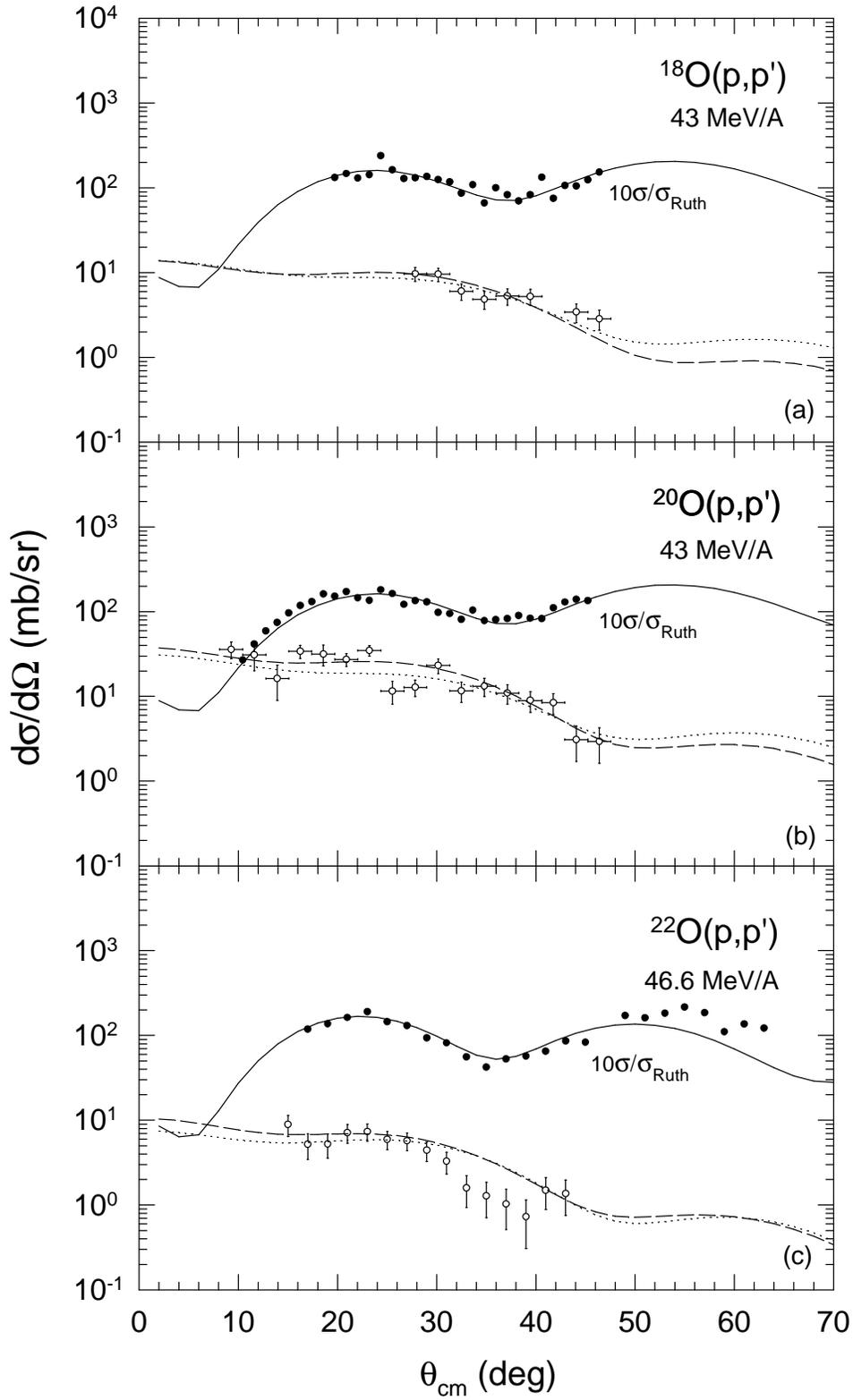}
\caption {Same as Fig. 2 but elastic cross sections are plotted
as ratio to Rutherford cross sections (scaled by a factor of 10
for clarity). The dashed (Legendre expansion method) and dotted
(deformed optical potentials) lines correspond to coupled
channel calculations of the inelastic cross sections using two 
different collective form factors.}
\end{figure}

\begin{figure}[h]
\includegraphics[width = 0.70\hsize,clip]{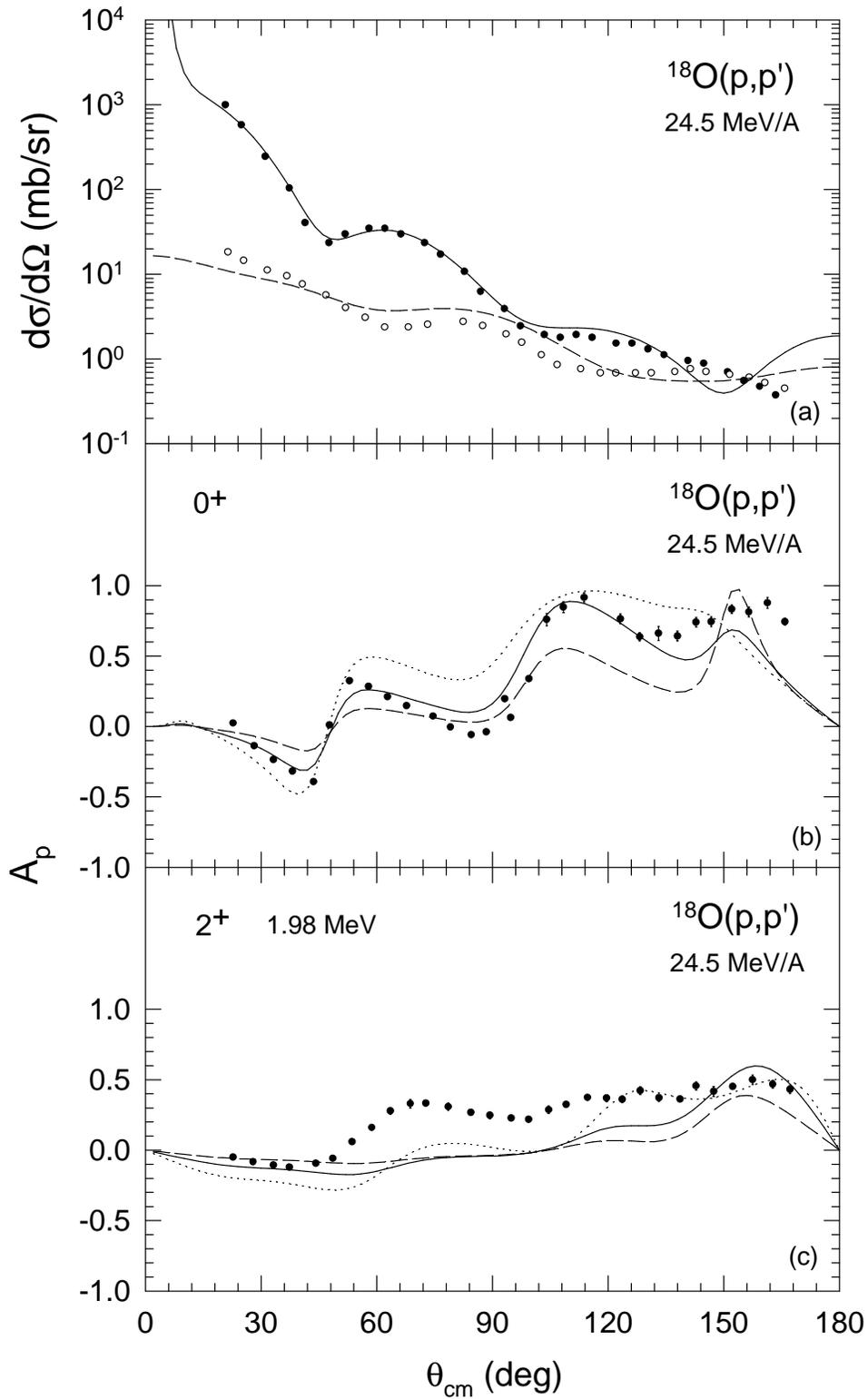}
\caption{The experimental angular distributions and folding model
calculations of p + $^{18}$O at 24.5A MeV~\cite{ES74} for (a)
elastic and inelastic [$E^*$ = 1.98 MeV (2$^+$)] differential cross
section, (b) analyzing power for the ground state, (c) analyzing
power for the excited state [$E^*$ = 1.98 MeV (2$^+$)]. The QRPA
transition densities are used in the calculations. In (b) and (c)
the continuous, dashed and dotted lines correspond to calculations
for 1.0, 0.5 and 2.0 times the spin-orbit term respectively.}
\end{figure}

\begin{figure}[h]
\includegraphics[width = 0.70\hsize,clip]{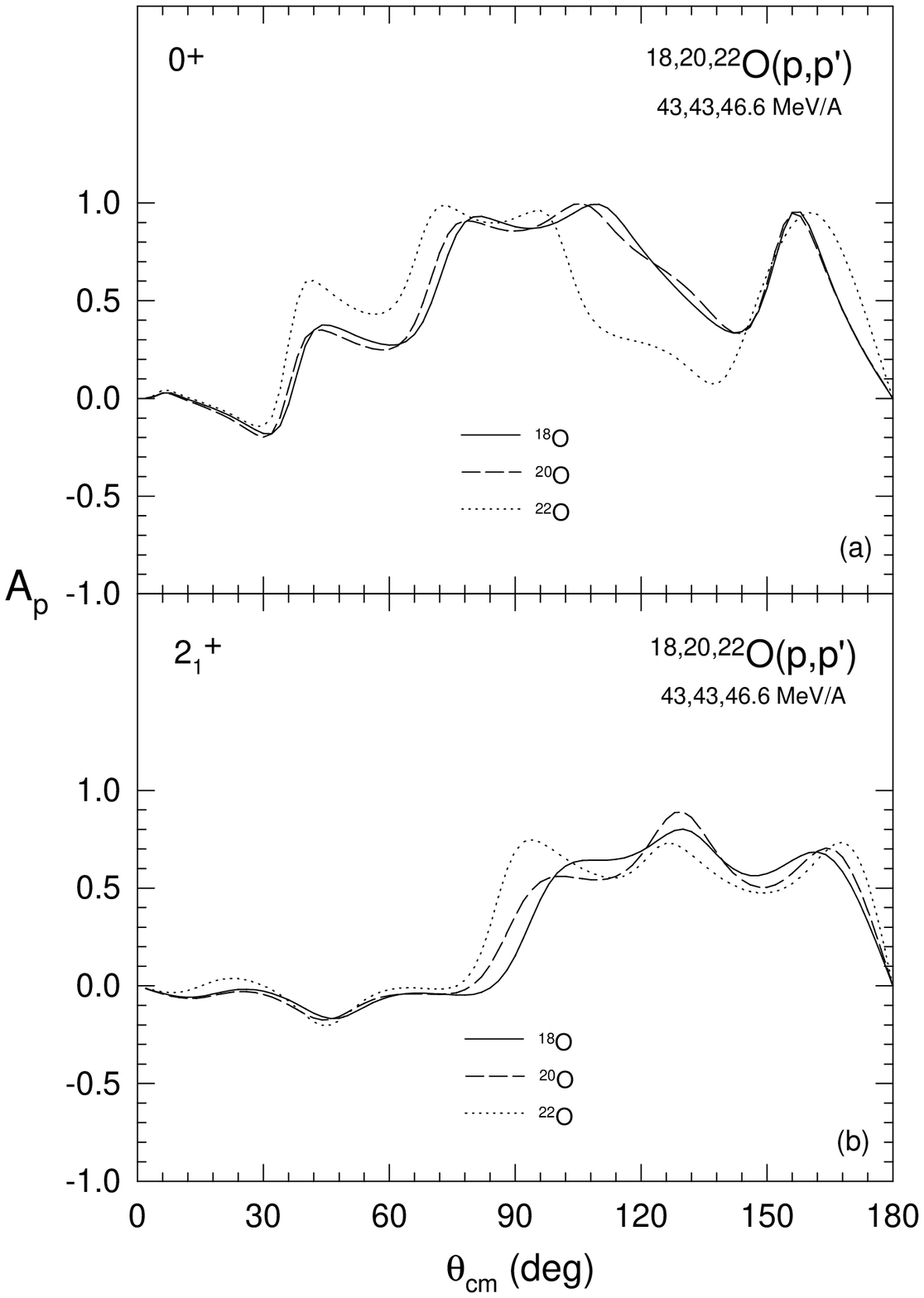}
\caption {The folding model calculations of p + $^{18}$O at 43A MeV
[$E^*$ = 1.98 MeV (2$^+$)], p + $^{20}$O at 43A MeV [$E^*$ = 1.67
MeV~(2$^+$)], p + $^{22}$O at 46.6A MeV [$E^*$ = 3.17 MeV (2$^+$)]
for (a) analyzing power for the ground state, (b) analyzing power
for the excited state. The QRPA transition densities are used in the
calculations.}
\end{figure}


\newpage
\begin{references}

\bibitem{TA97} O. Tarasov et al.,Phys. Lett. B 409 (1997) 64
\bibitem{SA99} H. Sakurai et al.,Phys. Lett. B 448 (1999) 180
\bibitem{BE06} E. Becheva et al., Phys. Rev. Lett. 96 (2006) 012501; see
references therein
\bibitem{JE99} J.K. Jewell et al.,Phys. Lett. B 454 (1999) 181
\bibitem{AM00} K. Amos, P. J. Dortmans, H. V. von Geramb, S. Karataglidis,
Adv. in Nucl. Phys. 25 (2000) 275
\bibitem{GU05} D. Gupta, D.N. Basu, Nucl. Phys. A 748 (2005) 402;
see references therein
\bibitem{GU02} D. Gupta, C. Samanta, Jour. Phys. G: Nucl. Part. Phys. 28 (2002) 85
\bibitem{KH03} D. T. Khoa, Phys. Rev. C 68 (2003) 011601(R)
\bibitem{KH00} E. Khan et al., Phys. Lett. B 490 (2000) 45
\bibitem{ES74} J. L. Escudie, R. Lombard, M. Pignanelli, F. Resmini and A. Tarrats,
Phys. Rev. C 10 (1974), 1645
\bibitem{GU00} D. Gupta, C. Samanta, R. Kanungo, Nucl. Phys. A 674 (2000) 77
\bibitem{RE74} H. Rebel, G. Hauser, G. W. Schweimer, G. Nowicki, W. Wiesner
and D. Hartmann, Nucl. Phys. A 218 (1974) 13
\bibitem{PE93} F. Petrovich, S. K. Yoon, M. J. Threapleton, R. J. Philpott,
J. A. Carr, Nucl. Phys. A 563 (1993) 387
\bibitem{BA04} D. N. Basu, Jour. Phys. G: Nucl. Part. Phys. 30 (2004) B7
\bibitem{SA79} G. R. Satchler and W. G. Love, Phys. Rep. 55 (1979) 183
\bibitem{r9}  G. Bertsch, J. Borysowicz, H. McManus and W.G. Love, Nucl. Phys. A 284 (1977) 399
\bibitem{r10} A.M. Kobos, B.A. Brown, R. Lindsay and G.R. Satchler, Nucl. Phys. A 425 (1984) 205
\bibitem{r11} A.K. Chaudhuri, Nucl. Phys. A 449 (1986) 243
\bibitem{KH02} E. Khan, N. Sandulescu, M. Grasso, N. Van Giai, Phys. Rev. C 66 (2002)
024309
\bibitem{DWUCK4} P. D. Kunz, DWUCK4, CHUCK3 documentation available online at
http://spot.colorado.edu/$^{\sim}$kunz/DWBA.html
\bibitem{RA01} S. Raman, C.W. Nestor Jr. and P. Tikkanen, Atomic Data and Nuclear Data Tables
78 (2001) 1
\bibitem{TH00} P. G. Thirolf et al., Phys. Lett. B 485 (2000) 16
\bibitem{KH01} E. Khan et al., Nucl. Phys. A 694 (2001) 103
\bibitem{KH002} D. T. Khoa et al, Nucl. Phys. A 706 (2002) 61

\end{references}
\end{document}